\documentclass[3p]{elsarticle}

\usepackage{graphicx}
\usepackage{amsmath}
\usepackage{amssymb}

\DeclareMathOperator{\erfc}{erfc}
\DeclareMathOperator{\erf}{erf}

\journal{Physics Letters A}

\begin{document}
\begin{frontmatter}

\title{Brownian diode: Molecular motor based on a semi-permeable Brownian particle with internal
  potential drop}
\author{A.V. Plyukhin}
%\email{aplyukhin@anselm.edu}
%\affiliation{ Department of Mathematics,
%Saint Anselm College, Manchester, New Hampshire 03102, USA 
%}
\address{
Department of Mathematics, Saint Anselm College, Manchester, New Hampshire 03102, USA 
}


\begin{abstract}
A model of an autonomous isothermal
Brownian motor with an internal propulsion 
mechanism  is considered. 
The motor is a Brownian particle which is semi-transparent
for molecules of surrounding ideal gas.  
Molecular passage through the particle is controlled by
a potential similar to that in the transition 
rate  theory, i.e. characterized by two stationary states with a finite energy
difference  separated by a potential barrier. The internal potential drop 
maintains  the diode-like asymmetry of molecular fluxes through the particle,
which results in the particle's stationary drift.  
%An analytic result for the stationary drift velocity, supported by
%a numerical simulation, is obtained for the regime of linear response. 

\end{abstract}

\end{frontmatter}
%\pacs{05.70.Ln, 05.40.-a, 05.20.-y}

%\maketitle
\section{Introduction}
Most natural (biological) molecular motors and  many 
artificial molecular machines are powered 
by chemical reactions.  Particular ways to convert chemical energy
to mechanical work vary considerably for different 
classes of motors. A number of   
prototypical models were suggested which are usually rather involved 
and require application of numerical methods~\cite{Doe,Ast,Rei,Liv,Kap}.
The purpose of this paper is to introduce a 
new simple model of autonomous Brownian motor 
%which allows a simple analytical treatment.
in which the complexity of involved chemical processes is 
hidden in a single parameter of an internal potential drop, 
not dissimilar to how  an electric battery is 
characterized by its voltage.

Our motor is  essentially a Brownian particle permeable for 
molecules of the surrounding thermal bath, which 
can flow in and out the particle. 
As a physical realization of permeable Brownian particles
one can mention for instance micro-gels~\cite{gels}, and 
lipid vesicles like liposomes~\cite{liposome}. 
When a molecule is inside 
the particle, it experiences a potential $U(x)$ similar to that
in the transition-state theory.
Calculations are particularly simple for a linear piecewise potential 
 $U(x)$ depicted  in  Fig. 1.
It is characterized, first, 
by the presence of a barrier which prevents the crossing of the particle by 
low-energy molecules,
and, secondly, by a nonzero potential difference $\Delta U$ between the 
two sides from the barrier.  
As discussed below, the internal propulsion force is generated by
the potential drop $\Delta U$ which, 
%is a built-in property of the motor.
similar to diode devices, results in asymmetry of molecular fluxes
through the particle and  a nonzero average momentum it 
receives in unit time.

%\enlargethispage{\baselineskip}

In the non-operational regime, when the average molecular flux through 
the particle is zero, 
a potential profile like that in Fig. 1 
can be formed naturally, for instance as  a result of 
asymmetric ion distribution across 
the particle.  The formation of nonzero potential difference between
the two sides of asymmetric lipid bilayers
was discussed in~\cite{Gurtovenko}.
However, it is evident that an external energy source 
is necessary to maintain an asymmetric charge distribution 
and internal potential drop
in the operational regime of nonzero flux across the particle
(otherwise the system would violate the second law of thermodynamics). 
%We believe that 
The function of a ``battery''
sustaining a stationary potential drop $\Delta U$ 
can be performed  by particle-bound ATP- or light driven
ion pumps, capable of moving ions against potential 
and/or concentration gradients (active transport)~\cite{Sancho,Ghosh}.
Certain proteins, 
%e. g., bacteriorhodopsin, 
which act as ion pumps 
in living cells,  are able to retain their function 
when 
%isolated from their native habitat and 
transplanted into  artificial 
membrane structures~\cite{Lanyi}. 
This appears to provide a technologically feasible way to synthesize
permeable Brownian objects with an internal potential drop.

\begin{figure}[htb]
\centerline{\includegraphics[height=3.1cm]{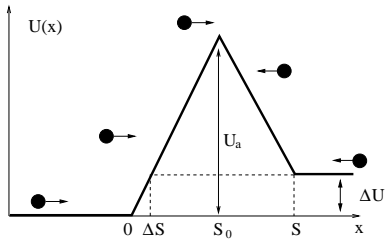}}
\caption{Potential energy $U(x)$ of molecules interacting with  
the permeable Brownian particle of size $S$. 
Coordinates $x=0$ and $x=S$ correspond to the left and right 
edges of the particle. Molecules (black circles) ``see'' the
barrier of height $U_a$ when coming from the left, 
and $U_a-\Delta U$ when coming from the right. 
For $\Delta U>0$, the particle drifts to the right.}   
\end{figure}

The model presented here is phenomenological in the sense that specific
mechanisms of active transport involved to sustain the potential drop 
$\Delta U$ are not identified. Of course, such an approach does not 
allow for the  determination of 
the motor's efficiency~\cite{Broeck}, but is sufficient to
evaluate transport characteristics of the motor as a function of $\Delta U$,
which is a goal of this paper.

\section{Model}
%Let us discuss the model in more detail. 
We consider a permeable 
Brownian particle of size $S$ and mass $M$ 
immersed in a thermal bath of ideal
gas molecules of mass $m\ll M$. 
The particle and molecules are constrained to move in one dimension,
molecular distribution far from the particle is homogeneous
with concentration $n$, and the velocity distribution of molecules is 
Maxwellian 
\begin{eqnarray}
f_M(v)=\frac{1}{v_T\sqrt{2\pi}}
\exp\left\{-\frac{1}{2}\left(\frac{v}{v_T}\right)^2\right\},
\label{Maxwell}
\end{eqnarray} 
where $v_T=1/\sqrt{m\beta}$ and $\beta=1/k_BT$. 
Molecules interact with the particle with a piecewise linear
potential $U(x)$ depicted in Fig. 1. The direct 
interaction of molecules to each other is assumed to be negligible. 
Suppose the particle's interior corresponds to the interval $0<x<S$, 
%$x=0$ and $x=S$ are positions  of  the left and right ends of the particle, 
with $x=S_0$ being the position of the barrier's maximum.
When inside the particle,  a molecule  experiences 
a constant force $f=|dU/dx|$ 
directed to the left on the left slope $0<x<S_0$, and to the right 
on the right slope $S_0<x<S$.
The right arm of the potential is shorter than the left one by
$\Delta S$,  which results in the difference of baseline potential values 
(potential drop) $\Delta U=f\Delta S$.

The idealized model potential like that in Fig.1 can be formed 
by two electric  double layers with opposite polarities, infinitely
extended in the $y$ and $z$ directions.
In a realistic case of the  particle of finite dimensions,
one has to take into account a nonzero field outside the particle
and, in general,  a discontinuity of the potential at the particle 
boundary~\cite{Parker}, 
but these would  not change the problem qualitatively.

It is intuitively clear that for given geometry 
the average force exerted on the particle by molecules
coming from the left exceeds the force from molecules
coming from the right, and thus the particle would drift to the right. 
Our aim is to find
the stationary drift velocity of the particle $\langle V\rangle$.       
(A more simple model with a monotonic step-like potential $U(x)$ shows
a nonzero drift, but no friction, and therefore, in our opinion, 
is of less interest.)

%It perhaps also should be stressed that the problem 
%we consider is essentially nonequilibrium.  
%Since molecular collisions are negligible 
%and the thermal bath bath is asssumed to be infinitely long,
%there is no mechanism
%to reach thermal equilibrium between the molecules from the left and right 
%sides of the particle, when their concentrations 
%would differ by the factor $\exp(-\beta \Delta U)$. 
%Instead, we consider the situation when 
%concentrations of $L$ and $R$-molecules bounding to
%the particle are the same and equal to the concentration in the
%unperturbed bath in the absence of the particle.
%Note that the situation is somewhat similar to the problem about 
%accelerating of ions of plasma in electric double layer~\cite{plasma}. 

In lowest order in the mass ratio parameter $m/M$,
which is sufficient for our purposes, 
one can justify the standard Langevin equation for the particle's velocity $V$,
\begin{eqnarray}
M\frac{dV(t)}{dt}=-\gamma \,V(t)+F(t),
\label{Langevin}
\end{eqnarray}
where however the 
fluctuating force $F(t)$ is not zero-centered, so
that the average stationary drift velocity is
\begin{eqnarray}
\langle V\rangle =\frac{\langle F\rangle}{\gamma}.
\label{drift1}
\end{eqnarray}
Even for the simple model outlined above, the analytic evaluation of 
$\langle F\rangle$ and $\gamma$ is rather involved except when 
the potential drop is small and the barrier is high,
\begin{eqnarray}
\beta\,\Delta  U \ll 1,\qquad \beta\, U_a \gtrsim 1.
\label{limits}
\end{eqnarray} 
The first condition allows one to evaluate  the average fluctuating force 
in a linear approximation in $\Delta U$, in which case the drift velocity 
can be written as
\begin{eqnarray}
\langle V\rangle=\sigma\Delta U,
\end{eqnarray} 
where the parameter $\sigma$ can be called the mobility.
The second condition, $\beta\, U_a \gtrsim 1$, implies that  
the friction coefficient $\gamma$ can be calculated 
in the limit of an infinitely high barrier,  
when the fluctuating force $F(t)$ is zero-centered, and 
the standard fluctuation-dissipation relation  holds 
\begin{eqnarray}
\gamma=\beta\int_0^\infty dt\,\langle
F(0)F(t)\rangle.
\label{fdt}
\end{eqnarray}

From a practical point of view the conditions (\ref{limits}) are of less
interest since the corresponding mobility $\sigma$ is very small. 
Yet in what follows we evaluate $\sigma$ assuming for the sake of simplicity
that conditions (\ref{limits}) are satisfied.
One naturally expects that in this case  
the temperature dependence   
is given by an Arrhenius law $\sigma=\sigma_0\exp(-\beta U_a)$, so that 
our major
interest will be in the pre-exponential factor $\sigma_0$.

\section{Impermeable particle}
In this section, we evaluate the correlation function of the Langevin
fluctuating force $F(t)$ and,  
using the fluctuation-dissipation relation (\ref{fdt}),
the friction coefficient $\gamma$, 
for the case of a high barrier or low temperature, 
when the molecular barrier crossing is  negligible. 
Clearly, in this case of an impermeable particle the potential drop does
not play any physical role and can be set to zero, $\Delta U=0$.
Recall that we need $\gamma$ in order to find the drift velocity of the motor 
with Eq.(\ref{drift1}).
As  will be found below, for the given model
\begin{eqnarray}
\gamma=
4\sqrt{\frac{2}{\pi}}\,n\,m\,v_T,
\label{gamma_0}
\end{eqnarray}
where $n$ is the concentration of bath molecules.
Equipped with this result, 
the reader not interested in derivation details may skip the rest of 
this section. 
Yet, since our evaluation is
microscopic and  exact, 
it is also a problem of interest of its own. It is also  not entirely 
distracting since in the next section a similar method will be applied to 
evaluate $\langle F\rangle$  for a permeable particle.

It is convenient to write 
the fluctuating force as the sum of contributions 
from molecules coming from the left 
and right, $F=F_L+F_R$. 
For the given case $F$ is zero centered, $\langle F\rangle=0$, 
and the statistical properties of
$F_L$ and $F_R$ are the same.  Then it is sufficient to consider only one, 
say the ``left'', contribution, since  the correlation function  of the total force 
can be written as 
\begin{eqnarray}
\langle F(0)F(t)\rangle=
2\langle\!\langle F_L(0)F_L(t)\rangle\!\rangle,
\end{eqnarray}
where double brackets denote the cumulant
$\langle\!\langle A\,B\rangle\!\rangle=
\langle A\,B\rangle-\langle A\rangle\,\langle B\rangle$.

\begin{figure}[htb]
\centerline{\includegraphics[height=3.2cm]{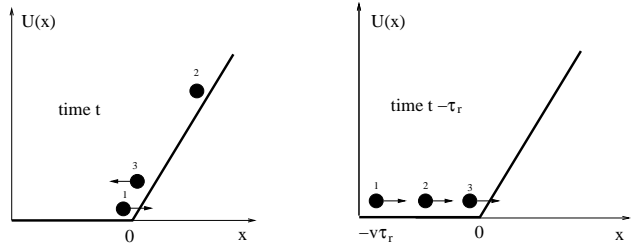}}
\caption{Left figure: 
a snapshot at time $t$ of three $L$-molecules with the same 
initial velocity $v$ which are inside the particle ($x>0$).
Arrows show velocity directions.  
Molecule $1$ has just entered the
particle, molecule $2$ has  reached a maximal position, molecule
$3$ is about to leave the particle's interior.  Right figure: a snapshot 
of the same molecules at an earlier time $t-\tau_r$. All three molecules are
outside the particle and occupy the interval $-v\tau_r <x<0$. }
\end{figure}

As known from the microscopic theory of Brownian motion~\cite{Mazo}, 
to  lowest order in the small mass ratio parameter $m/M$ 
the fluctuation 
force $F(t)$ can be evaluated neglecting the particle's motion, 
i.e. as a pressure on an infinitely heavy particle.
Suppose the particle's left edge is fixed at $x=0$, so that
$L$-molecules (coming from the left) 
experience the potential $U(x)$ which equals $fx$ 
for $x>0$ and zero for $x<0$, see Fig. 2. 
Each $L$-molecule inside the particle ($x>0$) gives
the same contribution $f$ to the total force on the particle, which therefore 
can be written as
\begin{eqnarray}
F_L(t)=f\int_{-\infty}^\infty \!\!\!dv\int_0^\infty \!\!\!dx\, N(x,v,t),
\label{force1}
\end{eqnarray}
where
\begin{eqnarray}
N(x,v,t)=\sum_{i}\delta(x-x_i)\delta(v-v_i)
\label{density}
\end{eqnarray}
is the microscopic phase density of $L$ molecules.

Let us define the residence time $\tau_r(v)$ as the time which 
an $L$-molecule with 
initial velocity $v$ spends inside the particle, i.e. in the region $x>0$. 
Suppose a molecule enters the particle at $t=0$ with the initial velocity
$v$.  Inside the particle the velocity changes with time as $v(t)=v-(f/m)t$. 
At the moment when the molecule leaves the particle,  its velocity is $-v$. 
Then the equation $v-(f/m)\tau_r=-v$ gives the residence time 
\begin{eqnarray}
\tau_r(v)=\frac{2mv}{f}. 
\label{residence_time}
\end{eqnarray}
For brevity, we shall often suppress the argument $v$ 
of the function $\tau_r(v)$.

In order to evaluate the correlation function of the force $F_L(t)$, 
it is convenient to express it in terms of phase density  
at an earlier time $t-\tau_r$. All molecules,
which at time $t$ are inside the particle $x>0$, at time $t-\tau_r$ were
outside in the region $-v\tau_r<x<0$, see Fig. 2. 
Then, instead of (\ref{force1}), one can write
\begin{eqnarray}
F_L(t)=f\int_{0}^\infty \!\!\!dv\int_{-v\,\tau_r}^0 \!\!\!dx\,N(x,v,t-\tau_r).
\label{force2}
\end{eqnarray}
The advantage of this form is that it refers to molecules 
in the potential free region $x<0$ where the 
velocities of molecules are constant, so that the following relation holds
\begin{eqnarray}
N(x,v,t)=N(x-vt, v, 0) \equiv N_0(x-vt, v),
\label{trick}
\end{eqnarray}
where the notation $N_0(x,v)$ is introduced for the density at time $t=0$.
This relation allows one to reduce temporal correlations to single-time 
correlation in phase space. For an ideal gas the latter takes the form  
\begin{eqnarray}
\langle\!\langle N_0(x,v) \, N_0(x',v')\rangle\!\rangle=
n\,f_M(v)\,\delta(v-v')\,\delta(x-x'),
\label{cum1}
\end{eqnarray}
where $f_M(v)$ is the Maxwellian distribution (\ref{Maxwell}), and $n$ is the 
density of the gas.

With the above relations,  
the evaluation of the correlation function for the force 
$F_L(t)$ can be done as  follows. 
From (\ref{force2}), one gets
\begin{eqnarray}
\langle\!\langle F_L(t)F_L(0)\rangle\!\rangle=
f^2\int_{0}^\infty \!\!dv\int_{0}^\infty \!\!dv'
\int_{-v\,\tau_r}^0 \!\! dx
\int_{-v'\,\tau_r'}^0 \!\! dx'\,
\langle\!\langle N(x,v,t-\tau_r) N(x',v',-\tau_r')
\rangle\!\rangle,
\label{cum2}
\end{eqnarray}
where $\tau_r'=\tau_r(v')$. 
Using (\ref{trick}), we can write 
the cumulant of the phase density which appears here as
\begin{eqnarray}
\langle\!\langle N_0(x-tv+\tau_rv,v) \, N_0(x'+\tau_r'v',v')
\rangle\!\rangle.
\label{cum3}
\end{eqnarray}
From (\ref{cum1})-(\ref{cum3}), after integration over
$v'$,
one obtains
\begin{eqnarray}
\langle\!\langle F_L(t)F_L(0)\rangle\!\rangle=
n\,f^2\int_{0}^\infty \!\! dv\,f_M(v)
\int_{-v\,\tau_r}^0 \!\!\!dx
\int_{-v\,\tau_r}^0 \!\!\!dx'\,\,
\delta(x-t\,v-x').
\label{cum4}
\end{eqnarray}
The integral of the delta function over $x'$ equals one when
\begin{eqnarray}
-v\,\tau_r<x-t\,v<0,
\label{aux1}
\end{eqnarray}
and zero otherwise.
Since Eq.(\ref{cum4}) involves only negative $x$,  
the right side of the condition (\ref{aux1}) is
satisfied (we assume $t>0$), 
while the left side requires $x>-v(\tau_r-t)$.
Then Eq. (\ref{cum4}) takes the form
\begin{eqnarray}
\langle\!\langle F_L(t)F_L(0)\rangle\!\rangle=
n\,f^2\int_{0}^\infty \!\! dv\,f_M(v)
\int_{-v\,\tau_r}^0 \!\!\!\!\!\ dx\,\theta(x+v(\tau_r-t))=
n\,f^2\!\!\int_{0}^\infty \!\!\!\!dv\,f_M(v)\,v\, (\tau_r-t)\,\theta(\tau_r-t),
\label{cum5}
\end{eqnarray}
where $\theta(x)$ is the Heaviside step-function.
Recalling expression
(\ref{residence_time}) for the residence time,  one obtains
\begin{eqnarray}
\langle\!\langle F_L(t)F_L(0)\rangle\!\rangle\!=\!
2 nmf\!\!\!\int\limits_{ft/2m}^\infty \!\!\!dv\,f_M(v)\,
\left(
v^2-\frac{ft}{2m}\,v
\right),
\end{eqnarray}
which eventually gives
\begin{eqnarray}
\langle F(t)F(0)\rangle=
2\langle\!\langle F_L(t)F_L(0)\rangle\!\rangle
=2n\,m\,f\,v_T^2\,\erfc\left(\frac{t}{t_0}\right),
\end{eqnarray}
where $\erfc(x)=1-\erf(x)$ is the complementary error function, and
the characteristic time of correlation is 
$t_0=\sqrt{2}\,\tau_r(v_T)=2\sqrt{2}m\,v_T/f$.
Then for the friction coefficient $\gamma$ we obtain the 
result (\ref{gamma_0}),  
\begin{eqnarray}
\gamma=\beta\!\int_0^\infty \!\!dt\, \langle F(t)F(0)\rangle=
4\sqrt{\frac{2}{\pi}}\,n\,m\,v_T.
\label{gamma}
%\nonumber
\end{eqnarray}
%Our numerical simulation confirms this result,  
%showing  exponential relaxation of the average velocity
%of the particle $\langle V(t)\rangle=V(0)\exp(-\gamma t)$
%with $\gamma$ very close to (\ref{gamma}). 
%As discussed in the next section, 
%the above analysis can be readily extended to the case of the potential of
%finite height $U_a$ and non-negligible molecular barrier crossings. 
%However, an analytical 
%expression for the correlation $\langle F(t)F(0)\rangle$
%is getting difficult to obtain, and it is more natural 
%to use numerical simulation to evaluate
%$\gamma$ for this case. 
%We postpone this issue till the last section.  
It is interesting to note that 
this expression,
which we found here for a linear potential $U(x)$ (constant repulsive force), 
coincides with $\gamma$  for
a similar model where molecules interact with the particle 
with a parabolic potential (linear repulsive force)~\cite{PS}.
The same result for $\gamma$ can be obtained from 
the model with
instantaneous (and therefore binary) particle-molecule collisions~\cite{Kampen}.
For these models correlation functions 
$\langle F(0)F(t)\rangle$ are different, 
but the integral (\ref{gamma}) is the same.

%It is somewhat unexpected that the above expression (\ref{gamma}) for $\gamma$,
%which we found here for a linear potential $U(x)$ (constant repulsive force), 
%turns out to be the same as for
%a similar model where molecules interact with the particle 
%with a parabolic potential (linear repulsive force)~\cite{PS}. We can
%currently offer no explanation for the coincidence. 
%Note also that the result (\ref{gamma}) is four times larger 
%than for the model with
%instantaneous particle-molecule collisions~\cite{Dunkel}.  

\section{Permeable particle}
In this section we assume that the potential barrier height is large but 
finite, $\beta U_a\gtrsim 1$, so that hot molecules of the thermal bath 
can pass through the particle.  
For a non-zero internal potential drop, $\Delta U\ne 0$, 
the fluctuating force $F$ in the Langevin equation (\ref{Langevin}) 
is not zero-centered, which causes the particle's drift. 
Our goal is to evaluate 
$\langle F\rangle$ and the particle's drift velocity 
$\langle V\rangle=\langle F\rangle/\gamma$  (with $\gamma$ found in the previous section)  
within a linear approximation, 
treating $\Delta U$ as a small perturbation.

As in the previous section, we assume that the particle occupies the interval
$ 0<x<S$, and we write the force as the sum of contributions from molecules
coming from the left and from the right, $F=F_L+F_R$.  
Consider first $F_L$.   
It is convenient to analyze it as the sum of two parts,
\begin{eqnarray}
F_L=F_L^{(1)}+F_L^{(2)}.
\label{FL1}
\end{eqnarray}
The first part $F_L^{(1)}$ comes from ``slow'' molecules  
with initial velocities (before entering the particle) less than 
the activation velocity
\begin{eqnarray}
v_a=\sqrt{\frac{2U_a}{m}},
\label{va}
\end{eqnarray}
which is the  minimal speed for $L$-molecules to cross the barrier.
The contribution $F_L^{(2)}$ corresponds to ``fast'' molecules with 
initial velocities $v>v_a$.

\begin{figure}[htb]
\centerline{\includegraphics[height=2.8cm]{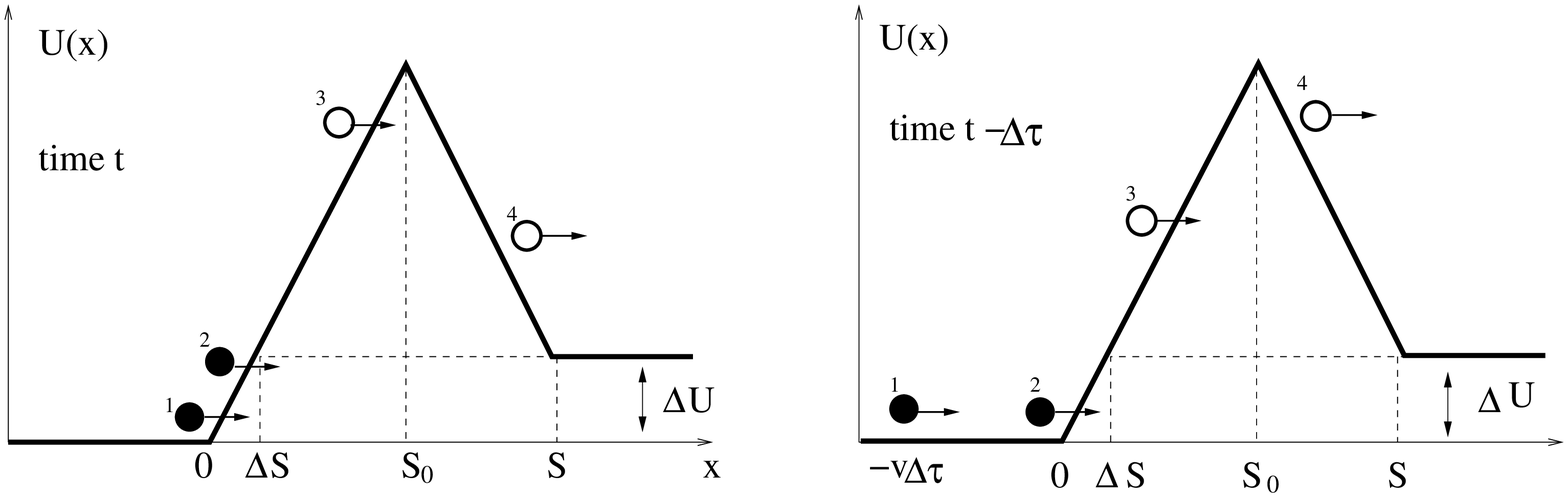}}
\caption{Left figure: a
snapshot of four fast $L$-molecules with the same initial velocity $v>v_a$.
Only molecules located in the interval $0<x<\Delta S$ 
(molecules $1$ and $2$) give nonzero contribution to the average force
$\langle F_L\rangle$ exerted on the particle by $L$-molecules. 
Contributions from molecules with $\Delta S<x<S_0$ 
(molecule $3$) and $S_0<x<S$ (molecule  $4$) cancel out.
Right figure: a snapshot 
of the same molecules at an earlier time $t-\Delta\tau$, when
molecules $1$ and $2$ were are outside the particle.}
\end{figure}

For molecules with $v<v_a$ the barrier is impenetrable, and in order 
to evaluate 
$F_L^{(1)}$ one can 
borrow the results of the previous section. Namely, 
as illustrated in Fig. 2,  
one can  
write $F_L^{(1)}$
at a given time $t$ in terms of the phase density of molecules  
at an earlier time $t-\tau_r$ as follows
\begin{eqnarray}
F_L^{(1)}(t)=f\int_0^{v_a} dv \int_{-v\tau_r}^{0} dx \,N(x,v,t-\tau_r).
\label{FL1_1}
\end{eqnarray}
Here the residence time $\tau_r(v)$ is given by (\ref{residence_time}),
and $f=|dU/dx|$. 
For the  potential-free region $\langle N(x,v,t)\rangle=n\,f_M(v)$, then 
from (\ref{FL1_1}) and (\ref{residence_time}) one  obtains
\begin{eqnarray}
\langle F_L^{(1)}\rangle=2\,m\,n\int_0^{v_a} \!\!dv\, f_M(v)\,v^2.
\label{FL1_2}
\end{eqnarray}
In the limit of impermeable particle, $v_a\to\infty$, the above equation gives
the result of the elementary kinetic theory for the pressure force $\langle
F\rangle=mn\langle v^2\rangle=n/\beta$.

Consider now the second part $F_L^{(2)}$ in Eq.(\ref{FL1}), which 
comes from fast molecules with
initial velocities  $v>v_a$.  Such molecules cross the barrier and 
can occupy both sides of the potential hill. 
Clearly, the average forces exerted by  fast $L$-molecules 
on the left and right slopes cancel out, except for the  molecules near the base
of the left slope with coordinates $0<x<\Delta S$, see Fig. 3. 
Let  $\Delta \tau$ be the time
for an $L$-molecule with initial velocity $v$ to climb to the 
altitude $\Delta U$,
\begin{eqnarray}
\Delta\tau=  \frac{mv}{f}\left\{
1-\sqrt{1-\frac{2\Delta U}{mv^2}}
\right\}\approx\frac{\Delta U}{f\,v}.
\end{eqnarray}
Similar to (\ref{FL1_1}), one can write
\begin{eqnarray}
F_L^{(2)}(t)=f\int_{v_a}^{\infty} dv \int_{-v\Delta\tau}^{0} dx \,
N(x,v,t-\Delta\tau),
\end{eqnarray}
which gives
\begin{eqnarray}
\langle F_L^{(2)}\rangle=n\Delta U\int_{v_a}^{\infty} dv \,f_M(v).
\label{FL2}
\end{eqnarray}
%Note that this  result
%does not depend on $f$, the value of the microscopic force of the
%particle-molecule interaction. 

The same method can be applied to evaluate two parts of the force 
\begin{eqnarray}
F_R=F_R^{(1)}+F_R^{(2)},
\end{eqnarray} 
exerted on the particle by $R$-molecules. 
The only difference is that  for molecules coming from the right 
the potential barrier is lower,  $U_a-\Delta U$, and so is  
the activation velocity
$\tilde v_a=[2(U_a-\Delta U)/m]^{1/2}$. 
In the linear approximation, 
\begin{eqnarray}
\tilde v_a\approx v_a-\Delta v_a,\qquad \Delta v_a=\frac{v_a \Delta U}{2U_a},
\label{Va_2}
\end{eqnarray}
where $v_a$ is the activation velocity for the $L$-molecules given 
by Eq.(\ref{va}). Using the same argument as for $L$-molecules, 
the contribution from slow $R$-molecules  with $|v|<\tilde v_a$,
which cannot cross the barrier,
can be written as 
\begin{eqnarray}
\langle F_R^{(1)}\rangle=-2\,m\,n\int_0^{\tilde v_a} \!\!dv\, f_M(v)\,v^2.
\end{eqnarray}
In the linear approximation this takes the form
\begin{eqnarray}
\langle F_R^{(1)}\rangle=-2\,m\,n\int_0^{v_a} \!\!dv\, f_M(v)\,v^2
+2mn\,f_M(v_a)v_a^2 \,\Delta v_a,
\nonumber
\end{eqnarray}
or
\begin{eqnarray}
\langle F_R^{(1)}\rangle=-\langle F_L^{(1)}\rangle
+2n\,f_M(v_a)v_a\,\Delta U.
\end{eqnarray}
The contribution from fast $R$-molecules with $|v|>\tilde v_a$ is similar 
to Eq. (\ref{FL2}) for fast $L$-molecules. In fact, in the linear 
approximation the
two expressions coincide,
\begin{eqnarray}
\langle F_R^{(2)}\rangle\!=\!n\Delta U\!\!\int_{\tilde v_a}^\infty \!\!dv\, f_M(v)\approx
n\Delta U\!\!\int_{v_a}^\infty \!\!dv\, f_M(v)\!=\!\langle F_L^{(2)}\rangle
\end{eqnarray}
Combining the above results,  the total average fluctuating force  
can be written as
\begin{eqnarray}
\langle F\rangle=\Bigl\{
2n\,f_M(v_a)v_a+2n\int_{v_a}^\infty \!\!dv\, f_M(v)
\Bigr\}\,\Delta U.
\end{eqnarray}
Explicitly this reads
\begin{eqnarray}
\langle F\rangle=n\,\phi\left(\frac{v_a}{v_T}\right)\,\Delta U,
\label{F_general}
\end{eqnarray}
where the dimensionless function $\phi(x)$ is 
\begin{eqnarray}
\phi(x)=\frac{2}{\sqrt{2\pi}}\, x\exp\left(-\frac{x^2}{2}\right)+\erfc\left(\frac{x}{\sqrt{2}}\right).
\label{exact}
\end{eqnarray}
Here the first and second terms originate from slow and fast molecules, 
respectively.
For large $x$ (high barrier/low temperature) 
the second term is smaller than the first one by the factor of
order $1/x^2$ and can be neglected. Then, since 
$v_a/v_{T}=(2\beta U_a)^{1/2}$,  
one gets
\begin{eqnarray}
\langle F\rangle=\frac{2}{\sqrt{\pi}}\,n\,\sqrt{\beta U_a}\, e^{-\beta U_a}\,
  \Delta U.
\end{eqnarray}
This and the expression (\ref{gamma_0}) for $\gamma$ give for 
the stationary drift velocity $\langle V\rangle =\langle F\rangle/\gamma$
of the motor the following result
\begin{eqnarray}
\langle V\rangle=\sigma\,\Delta U, \qquad \sigma=\frac{1}{2\sqrt{2}}\,\,
\beta\,\sqrt{\frac{U_a}{m}}\,e^{-\beta U_a}.
\label{result}
\end{eqnarray}
As expected, the temperature dependence of 
the mobility $\sigma$ is 
Arrhenius-like. The square-root dependence of pre-exponential factor
on the barrier's height $U_a$ is perhaps a less obvious result. 
The mobility $\sigma$ depends neither  on the mass of the
motor nor on the concentration of molecules of the thermal bath.

\begin{figure}[htb]
\centerline{\includegraphics[height=5.9cm]{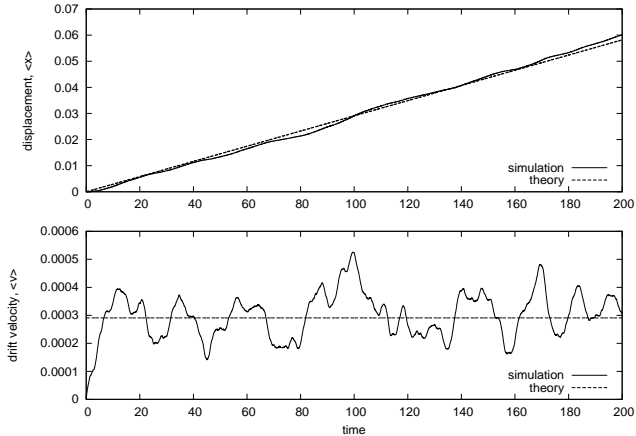}}
\caption{ Numerical simulation of the drift characteristics of the motor 
with $\beta U_a=5$, $\beta\Delta U=0.05$, and mass ratio $m/M=0.1$.
Upper figure: average displacement of the motor as a function of time. 
Lower figure: average drift velocity. Solid lines correspond to numerical
experiment data, averaged over $1.8\cdot 10^7$ trajectories. 
Dashed lines represent 
theoretical predictions given
by Eq.(\ref{result2}). Units of velocity, time, and displacement are, respectively, $v_0=v_T$, 
$\tau_0=\tau_r(v_T)/2$, and $x_0=v_0 \tau_0$.}
\end{figure}

\section{Discussion}

The autonomous mesoscopic  motor described in this paper   
is based on a permeable Brownian particle which propels due to
asymmetric transfer characteristics, and in this sense 
may be called a Brownian diode. 
%Such a motor does not challenge the second law of thermodynamics,
%because in the operational regime 
%(nonzero molecular flux)
%the internal potential drop across the particle  
%has to be maintained by outer means 
%(somewhat similar  situation
%takes place in formation    
%of the electric double layer in plasma~\cite{plasma}).
%which may involve complex biochemical processes.
The main result (\ref{result}) for the motor's mobility $\sigma$ 
is obtained under
asymptotic conditions of a high barrier and small potential drop 
(\ref{limits}). 
To verify this result, we performed a one-dimensional 
numerical simulation similar to
standard molecular dynamics with the velocity Verlet algorithm.
Instead of periodic boundary conditions 
that poorly fit the problem, a very large thermal bath  
with a Maxwellian velocity distribution (\ref{Maxwell}) and 
a constant density of molecules
was generated in the beginning of each simulation run.  
The size of the bath was of order
$c\,v_T \,t_s$, where $t_s$ is the duration of one simulation run, and
$c$ is a numerical parameter of order $10^2$.   
As far as the Brownian particle dynamics is concerned, 
the difference between a bath of such
size and the infinite one is that the particle immersed in the former 
may feel a deficit of hot molecules with velocities higher than $cv_T$.   
For $c\gtrsim 10^2$, it was found empirically that 
results do not depend on $c$, and thus 
the difference with the infinite bath is expected to be negligible. 
The assumption of non-interacting bath molecules strongly reduces
the amount of computations. It also allows some optimization by eliminating 
molecules with  initial velocity
directions away from the particle and  also those which 
are so slow and generated so far from the particle 
that cannot come to its proximity during the time of simulation. 
The ensemble average of the particle's coordinate and velocity 
as functions of time 
is taken by performing multiple runs, resetting the bath molecules 
with new initial conditions and averaging over the simulation runs.

Simulation  shows a good agreement with the prediction 
(\ref{result}) for small $\Delta U$ 
and when $\beta U_a\gtrsim 5$, see Fig.4. The agreement gets better
when one
takes into account the contribution of fast molecules given by the 
second term in Eq.(\ref{exact}). Since
$\erfc(x)\approx e^{-x^2}/\sqrt{\pi}x$ for large $x$, the  corrected expression
for the mobility takes the form
\begin{eqnarray}
\sigma=
\frac{1}{2\sqrt{2}}\,\,
\beta\,\sqrt{\frac{U_a}{m}}\,
\left(
1+\frac{1}{2\beta U_a}
\right)\,
\,e^{-\beta U_a}.
\label{result2}
\end{eqnarray}

\begin{figure}[htb]
\centerline{\includegraphics[height=5.9cm]{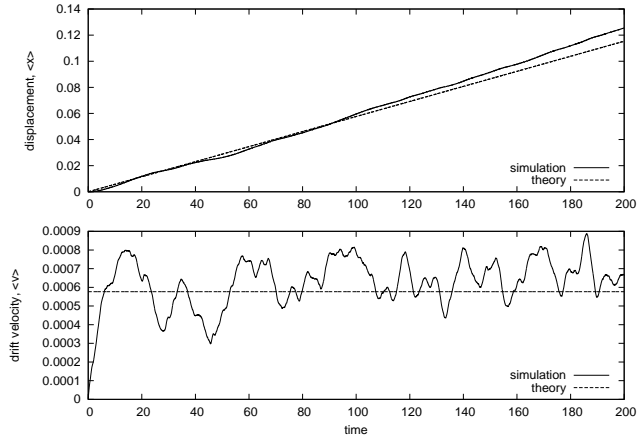}}
\caption{ 
Same as Fig. 4, but for a lower potential barrier $\beta U_a=4$.
For a lower barrier the disagreement of experimental and theoretical
results becomes more significant.
}
\end{figure}

For lower barrier heights,  $\beta U_a<5$,
the simulation  shows that Eqs.(\ref{result}) and (\ref{result2})
both noticeably  underestimate the mobility, see Fig. 5.
The discrepancy  comes from 
the  approximation  of an infinitely high barrier, 
which we have used to calculate
the friction  coefficient $\gamma$, Eq. (\ref{gamma}).  Clearly, 
this overestimates $\gamma$ and underestimates the drift velocity $\langle
V\rangle =\langle F\rangle/\gamma$. 
An analytical
evaluation   of $\gamma$ for the case of a finite barrier is feasible but 
somewhat cumbersome. One can instead find $\gamma$ from a numerical experiment
measuring velocity relaxation 
$\langle V(t)\rangle=V_0\exp(-\gamma t)$.
When one uses $\gamma$ found in this way, 
and $\langle F\rangle$ given by (\ref{F_general}),  
a good agreement with the numerical experiment may be restored. 
(This method is not very effective for the mass ratio $m/M\gtrsim 0.1$, 
when relaxation of the Brownian particle's velocity
is distinctly non-exponential.)  
 
%\enlargethispage{\baselineskip}

In our calculations, we assumed that the two slopes of the potential 
hill are equally steep. No qualitatively new feature appears 
if they are not (provided the potential is continuous). 
%It might be tempting to suggest that for 
%an asymmetric barrier 
%the particle may develop
%a finite drift velocity even if $\Delta U=0$. 
%This assumption contradicts the second law of
%thermodynamics and  incorrect. 
As shown above, the stationary mobility 
does not depend on  $U'(x)$.  
Respectively, 
for an asymmetric barrier with no potential drop, 
the drift may be only transient and vanishes 
in the long-time limit.  
This is confirmed by numerical simulation
for regimes within and beyond the linear response.
At this point, however, it perhaps should be recalled that 
in this study we ignored the particle's internal 
degrees of freedom. Taking them into account, one may expect 
that the model would reveal  new features
characteristic for granular Brownian motors, interacting with
bath molecules via inelastic collisions~\cite{granular}.

%This is also confirmed by numerical simulation
%for regimes beyond the linear response. 

%\section*{Acknowledgement}

%I thank Gregory Buck and the referees for their helpful comments. 

\bibliographystyle{elsarticle-num}

%\end{multicols}

\end{document}